# MEDICAL IMAGE DENOISING USING ADAPTIVE THRESHOLD BASED ON CONTOURLET TRANSFORM

S.Satheesh[1], Dr.KVSVR Prasad[2]

[1]Department of Electronics and Communication Engineering, G.Narayanamma Institute of Technology and Science, Hyderabad, India
*satheesh_ece@gnits.ac.in*
[2]Department of Electronics and Communication Engineering, D.M.S.S.V.H. College of Engineering, Machilipatnam, India
*kvsvr@yahoo.com*

*ABSTRACT*

*Image denoising has become an essential exercise in medical imaging especially the Magnetic Resonance Imaging (MRI). This paper proposes a medical image denoising algorithm using contourlet transform. Numerical results show that the proposed algorithm can obtained higher peak signal to noise ratio (PSNR) than wavelet based denoising algorithms using MR Images in the presence of AWGN.*

## 1. INTRODUCTION

Image denoising is a procedure in digital image processing aiming at the removal of noise, which may corrupt an image during its acquisition or transmission, while retaining its quality. Medical images obtained from MRI are the most common tool for diagnosis in Medical field. These images are often affected by random noise arising in the image acquisition process. The presence of noise not only produces undesirable visual quality but also lowers the visibility of low contrast objects. Noise removal is essential in medical imaging applications in order to enhance and recover fine details that may be hidden in the data.

## 2. CONTOURLET TRANSFORM

In the recent years, Do and Vetterli proposed a multiscale and multidirectional image representation method named contourlet transform [5, 6], which can effectively capture image edges and contours. The contourlet transform is constructed by Laplacian pyramid [4, 7, 10] (LP) and directional filter banks (DFB) [1, 2, 3, 9]. The Figure.1 illustrates the contourlet transformation, in which the input image consists of frequency components like LL (Low Low), LH (Low High), HL (High Low), and HH (High Low).

The Laplacian Pyramid at each level generates a Low pass output (LL) and a Band pass output (LH, HL, and HH). The Band pass output is then passed into Directional Filter Bank, which results in contourlet coefficients [8]. The Low pass output is again passed through the Laplacian Pyramid [7] to obtain more coefficients and this is done till the fine details of the image are obtained. Figure.2 shows the decomposition of brain MR Image.

Advanced Computing: An International Journal ( ACIJ ), Vol.2, No.2, March 2011

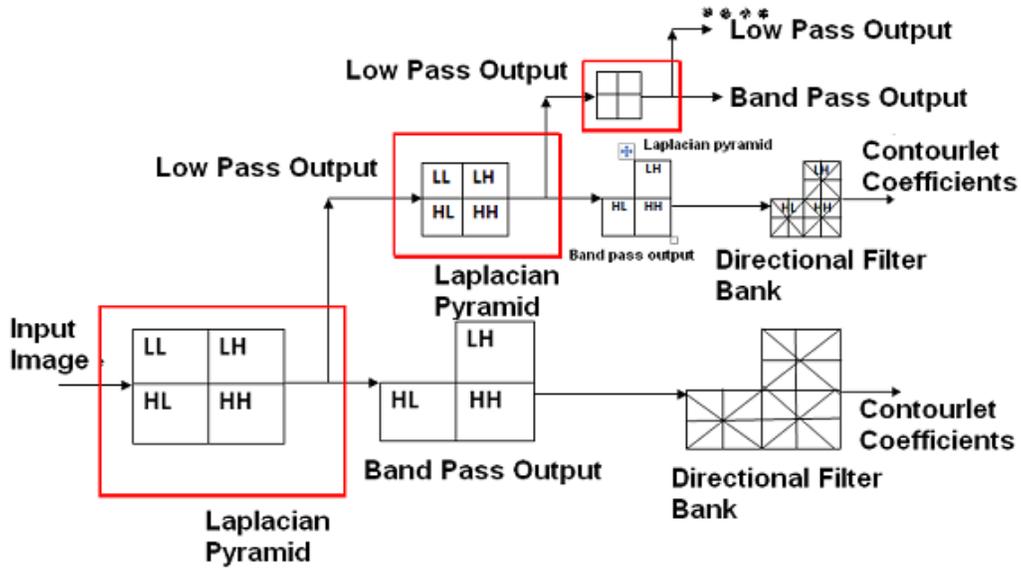

Figure1. Illustration of Contourlet Transform

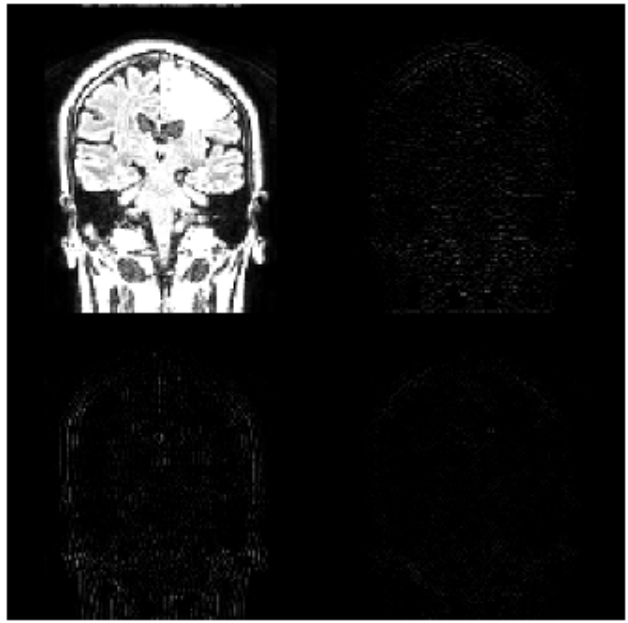

Figure2. Contourlet decomposition of brain MR Image

## 3. PROPOSED DENOISING ALGORITHM

A common approach for image denoising is to convert the noisy image into a transform domain such as the wavelet and contourlet domain, and then compare the transform coefficients with a fixed threshold. We propose an algorithm which defines a new threshold value to eliminate the corrupted pixels.

### 3.1 Estimation of Parameters

In this section we formulate the parameters which are utilized for denoising.

**3.1.1 Estimation of noise variance**

The noise variance is estimated using the mean absolute deviation (MAD) method and is given by

$$\sigma_n^2 = \left(\frac{median(c_{i,j})}{0.6745}\right)^2 \quad (1)$$

Where $c_{i,j}$ is the contourlet coefficients of noisy image

**3.1.2 Estimation of threshold**

The threshold T for the contourlet coefficients of noisy image is given by

$$T = \frac{3}{4} N\left(\frac{\sigma_n^2}{\sigma_g}\right) \quad (2)$$

Where $N$ is total number of pixels in the image and $\sigma_g$ is the standard deviation of the noisy image

**3.2 Algorithm Description**

The denoising method based on contourlet transform can be described as follows:

1. Perform contourlet transform to the noisy image; from the decomposition process the coefficients are extracted

2. Estimate the noise variance for each noisy image pixel using equation (1)

3. The threshold T for the contourlet coefficients of noisy image is calculated using equation (2)

4. If the contourlet coefficients are greater than the threshold, those coefficients are remained unchanged. If they are less, they are suppressed.

5. Then all the resultant coefficients are reconstructed by applying inverse contourlet transform, which results in denoised image.

## 4. RESULTS

In this section, simulation results are presented which is performed on the Spine and Brain MR images. Zero mean additive white Gaussian noise is added to the MR Images and denoised with wavelet based methods and the proposed method. The performance of the proposed method is compared with the hard threshold, soft threshold and Wiener filter in the wavelet domain using *PSNR*, which is defined as

$$PSNR = 10\log_{10}\frac{255^2}{MSE} \quad (3)$$

Where *MSE* denotes the mean square error for two $m \times n$ images $I(i, j)$ & $\hat{I}(i, j)$ where one of the images is considered a noisy approximation of the other and is given as

$$MSE = \frac{1}{mn} \sum_{i=0}^{m-1} \sum_{j=0}^{n-1} \left[ I(i, j) - \hat{I}(i, j) \right]^2 \quad (4)$$

For different Gaussian noise level densities we have obtained various PSNR values of Wavelet based methods and the proposed algorithm using contourlet transform, that are plotted. There is a significant improvement in PSNR values in the proposed algorithm. Figure.3 and Figure.4 show the comparative results among wavelet based methods and proposed algorithm using contourlet transform. From these quantitative results we infer that the new proposed algorithm using contourlet transform outperforms Wavelet based methods. This can be clearly seen from Figure.5 and Figure.6 that the background of the denoised images with contourlet transform appears smoother. The contourlet transform removes the noise pretty well in the smooth regions and also along the edges. So it is evident with the results that this proposed algorithm based on contourlet transform is best suit for Medical image applications.

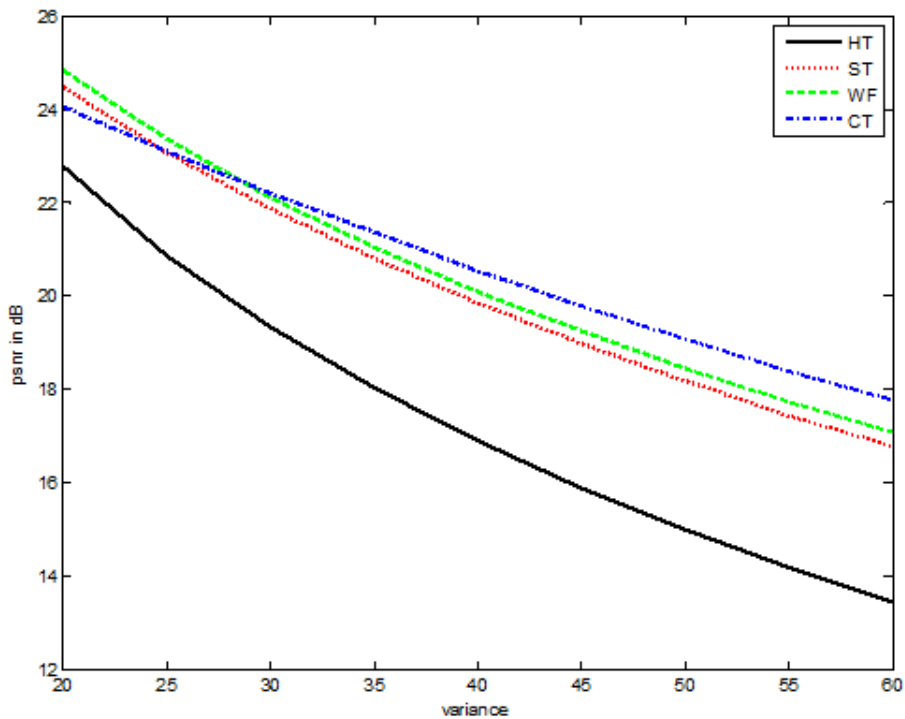

Figure3. The PSNR value of the de-noised Spine MR Image vs. Variance of the noise

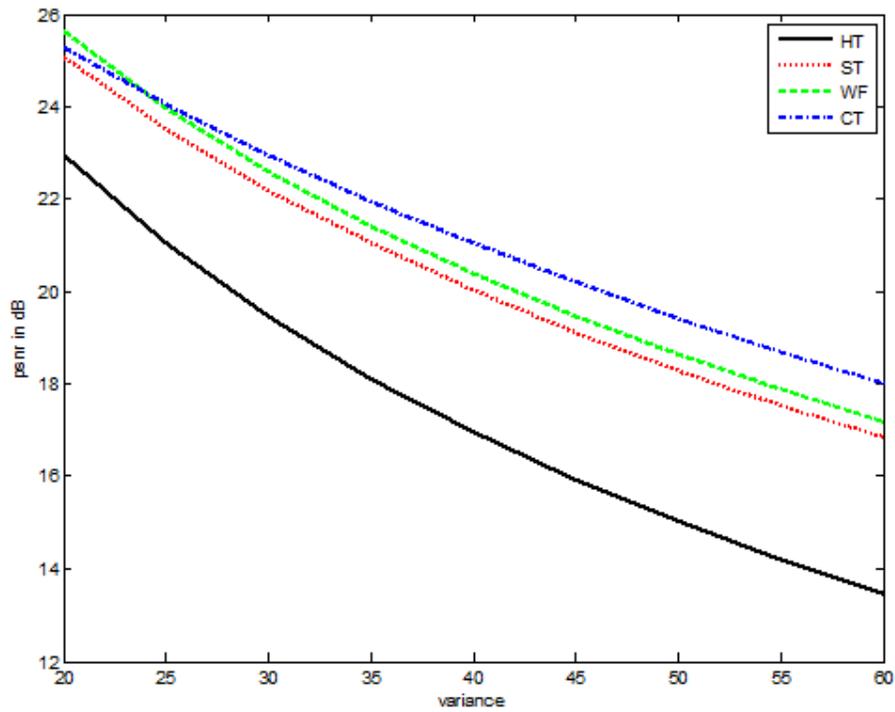

Figure4. The PSNR value of the de-noised Brain MR Image vs. Variance of the noise

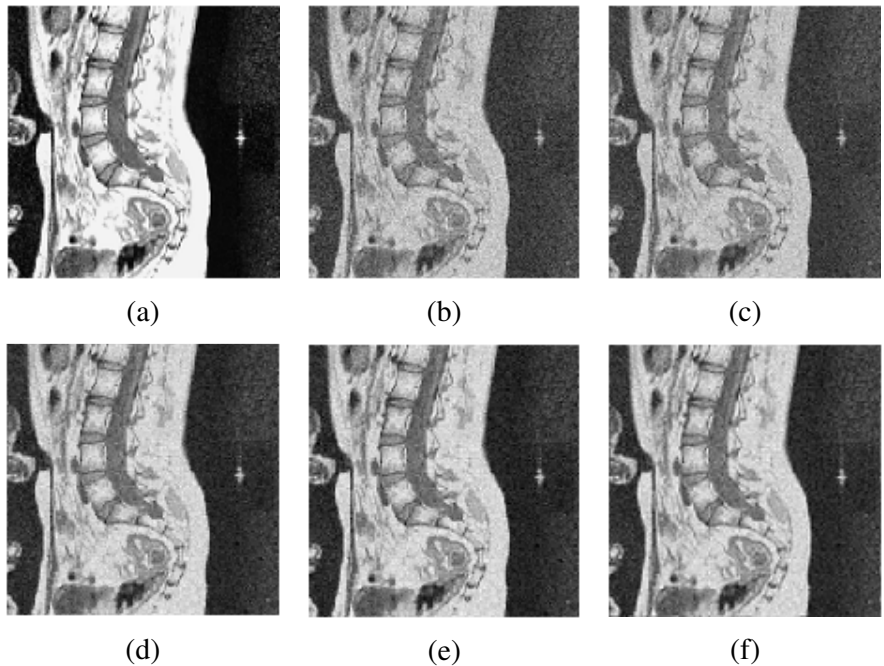

Figure5. Denoising of Spine MR Image for variance=25 (a) Original image (b)Noisy image (c)Denoised image with hard threshold (d) Denoised image with soft threshold (e) Denoised image with Wiener filter (f) Denoised image with contourlet transform

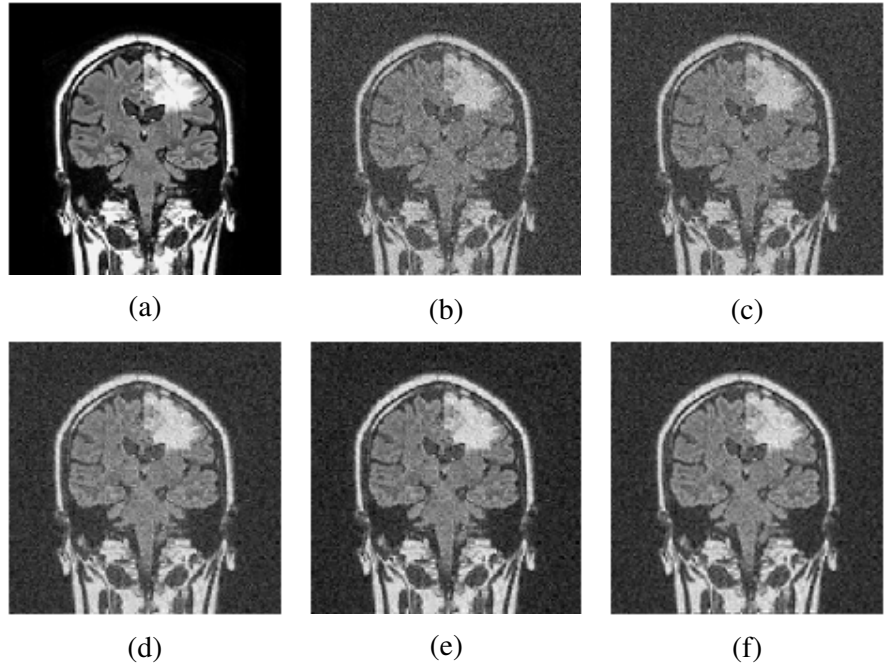

(a) (b) (c)

(d) (e) (f)

Figure6. Denoising of Brain MR Image for variance=30 (a) Original image (b)Noisy image (c)Denoised image with hard threshold (d) Denoised image with soft threshold (e) Denoised image with Wiener filter (f) Denoised image with contourlet transform

## 5. CONCLUSIONS

In this paper, the removal of Gaussian noise from MR Images has been discussed. Hence the new proposed algorithm based on the contourlet transform is found to be more efficient than the wavelet methods in Image Denoising particularly for the removal of Gaussian noise. Thus the obtained results in qualitative and quantitative analysis show that this proposed algorithm outperforms the wavelet methods both visually and in terms of PSNR.

## ACKNOWLEDGEMENTS

We wish to express our sincere thanks to Dr. K. Jitender Reddy, Consultant Radiologist, Dept. of Radiology & Imaging Sciences, Apollo Health City, Hyderabad for providing us with different MR Image datasets.

## REFERENCES

[1] Duncan D. Y. Po, Minh N. Do "Directional multiscale modeling of images using the Contourlet transform", IEEE Transactions on Image Processing, VOL. 15, NO. 6, Page No:1610-162, JUNE 2006

[2] Eslami,R., Radha,H., "The contourlet transform for image denoising using cycle spinning," presented at the Asimolar Conference on Signals, Systems and Computers, Pacific Grove, USA, vol.2, pp.1982-1986,November, 2003.

[3] Eslami, R. and Radha. H., "Translation-invariant contourlet transform and its application to image denoising," IEEE Transactions on Image Processing, Vol. 15, No.11, pp. 3362–3374, 2006

[4] M. N. Do and M. Vetterli, "Pyramidal directional filter banks and curvelets," Proc. IEEE Int. Conf. on Image Proc., vol. 3, 2001, pp. 158-161.

[5] M. N. Do and M. Vetterli, "Contourlets: a Directional Multiresolution image representation," Proceedings of 2002 IEEE International Conference on Image Processing, vol. 1,2002, pp 357-360.

[6] M. N. Do and M. Vetterli, "The Contourlet Transform: an efficient directional multiresolution image representation," IEEE Trans. On Imaging Processing, vol. 14(12), pp. 2091-2106, December 2005.

[7] P. J. Burt and E. H. Adelson, "The Laplacian pyramid as a compact image code," IEEE Trans. Commun., vol. 31, no. 4, 1983, pp. 532– 540.

[8] R.Sivakumar, G.Balaji,R.S.J.Ravikiran,R.Karikalan,S.Saraswathi Janaki "Image Denoising using Contourlet Transform" 2009 Second International Conference on Computer and Electrical Engineering,pp.22-24

[9] W.Y. Chan, N.F. Law, W.C. Siu, "Multiscale feature analysis using directional filter bank," Proceedings of the 2003 Joint Conference of the Fourth International Conference on Information, Communications and Signal Processing, December 2003, pp. 822-826.

[10] Zhiling Longa, b and Nicolas H. Younana "Denoising of images with Multiplicative Noise Corruption", 13th Europian Signal Processing Conference, 2005,,a1755.

## Authors

**S.Satheesh** received the B.Tech degree in Electronics and Communication Engineering from VRSEC (ANU, Vijayawada, India) in 2001, and the M.E (ECE) degree in Communication Engineering specialization from CBIT (OU, Hyderabad, India) in 2005.He is currently pursuing the Ph.D. degree under the guidance of Dr. KVSVR Prasad at Jawaharlal Nehru Technological University Hyderabad, India. He is the member of IEEE, ISTE, IAENG and IACSIT. His research interests are in the domain of Medical Image Processing and Signal Processing.

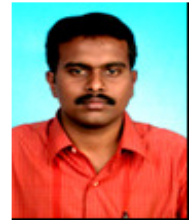

**Dr. KVSVR Prasad** obtained B Sc. Degree in 1963 from Andhra University, B.E (Telecommunication Engineering) in 1967 and M.E (ECE) Microwave Engineering specialization in 1977 from Osmania University. He received the Ph.D. in 1985 from IIT Kharagpur in strip and micro strip transmission lines. He published six papers in IEEE Transactions in MTT, Antenna and Propagation and EMI/EMC and three papers in National Conferences. He is fellow of IETE (life member). He worked in various capacities in the Department of ECE, Osmania University, Hyderabad. Presently he is working as professor and head, Department of ECE, D.M.S.S.V.H. college of engineering, Machilipatnam, India.

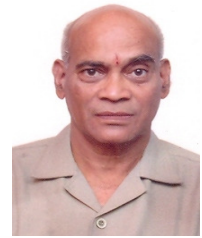